\documentstyle[aps
,epsfig%
,preprint%
]{revtex}
\newcommand{\nn}{\nonumber \\}
\newcommand{\Bx}{\bbox{x}}
\newcommand{\BA}{\bbox{A}}
\newcommand{\BP}{\bbox{P}}
\newcommand{\Bq}{\bbox{q}}
\newcommand{\Bk}{\bbox{k}}
\newcommand{\BJ}{\bbox{J}}
\newcommand{\Bv}{\bbox{v}}
\newcommand{\Br}{\bbox{r}}
\newcommand{\Bnabla}{\bbox{\nabla}}
\newcommand{\Neq}{\mbox{$\,/\!\!\!\!\!=\,$}}
\newcommand{\delu}[1]{\frac{\partial_{#1} U}{U}}
\begin{document}
\draft
\preprint{Guchi-TP-006}
\date{\today
}
\title{
Effective field theory of slowly-moving ``extreme black holes''
}
\author{Yoshitaka~Degura%
\thanks{Email address: {\tt c1997@sty.cc.yamaguchi-u.ac.jp}
}}
\address{Graduate School of Science and Engineering, 
Yamaguchi University, 
Yoshida, Yamaguchi-shi, Yamaguchi 753-8512, Japan
}
\author{Kiyoshi~Shiraishi%
\thanks{Email address: {\tt shiraish@sci.yamaguchi-u.ac.jp}
}}
\address{Graduate School of Science and Engineering, 
Yamaguchi University, 
Yoshida, Yamaguchi-shi, Yamaguchi 753-8512, Japan\\
and Faculty of Science, Yamaguchi University,
Yoshida, Yamaguchi-shi, Yamaguchi 753-8512, Japan
}
\maketitle
\begin{abstract}
We consider the non-relativistic effective field theory of
``extreme black holes'' in the Einstein-Maxwell-dilaton theory with
an arbitrary dilaton coupling. We investigate finite-temperature behavior
of gas of  ``extreme black holes'' using the effective theory.
The total energy of the classical many-body system is also derived.
\end{abstract}


\section{introduction}

Recently the properties of black holes in the Einstein-Maxwell-dilaton (EMD)
theory have been studied by many authors\cite{GiMa,GHS}. We are interested
in the EMD theory since it can be regarded as an
effective field theory of string and/or the gravitational
theory with a dimensional reduction. An ``extreme black hole'', which we
consider in the present paper, is a kind of solitons with balanced
three long-range forces (the gravitational force, the electric force, and
the dilatonic force). 
\footnote{If a dilaton field is coupled, the extreme limit of the black
hole solution has a singularity in general. Therefore, we use the
``quotation marks'' around the extreme black hole in this paper.}
For the balance, the charge and mass of the ``black hole'' must satisfy the
certain condition.

For the static case, the system with an arbitrary number of ``extreme black
holes'' is known to be described by the Papapetrou-Majumdar
solution\cite{Papa,Madj} and its descendants\cite{Myer,Shir}. In the
present paper, we consider slowly-moving ``extreme black holes''. Since
they can be regarded as point particles, 
we are able to describe the collective behavior of them
by a scalar field. Though the similar system has been
studied before\cite{FeEa,TrFe,Shi1,Shi2}, our approach is different from
previous works and may give new insights of the collective phenomena of such
objects. The paper is arranged as follows.

In Sec.~\ref{sec:2}, we derive a
theoretical model of a non-relastivistic scalar field incorporating the
low-energy interaction among ``extreme black holes''. As an application, we
examine an isothermal sphere of ``extreme black holes'' by using the
technique of finite-temperature field theory in Sec.~\ref{sec:3}. As a
result, we will show that the gas of ``extreme black holes'' lumps at
high temperature. In Sec.~\ref{sec:4}, we derive the expression for
the energy of the system after eliminating the potentials by the field
equations. We also obtain the total energy of the classical many-body
system. Section~\ref{sec:5} is devoted to conclusion. 

The derivation of the effective lagrangian for the model in
$(N+1)$ dimensions is exhibited in Appendix.

\section{Effective Lagrangian}
\label{sec:2}

In this section, we derive an effective lagrangian for 
``extreme black holes'' of which a typical velocity $v$ is small. 
In the classical theory, the action of particles with mass 
$m$ and charge $e_0$, coupled to the gravitational field, the 
electromagnetic field $A_{\mu}$, and the dilaton field $\phi$, is written as
\begin{equation}
I = \,- \int\!\! ds\:\Bigl[m\,e^{a\phi}+e_0 A_\mu \frac{dx^\mu}{ds}
\Bigr]\, ,
\label{eq:001}
\end{equation} 
where $a$ is the coupling constant for the dilaton field. Then, the
four-momentum of the particle is
\begin{equation}
 P_{\mu} = m e^{a\phi} g_{\mu\nu} \frac{dx^{\nu}}{ds} - e_0 A_{\mu}\, .
\label{eq:002}
\end{equation}
This four-momentum satisfies the following equation:
\begin{equation}
 g^{\mu\nu}(P_{\mu}+e_0 A_{\mu})(P_{\nu}+e_0 A_{\nu})+m^2 e^{2a\phi}=0\, .
\label{eq:003}
\end{equation}

In quantum theory, we can rewrite this equation (\ref{eq:003}) 
into a wave equation for a wave function $\varphi$:
\begin{equation}
 \Bigl[\,g^{\mu\nu}(P_{\mu} + e_0 A_{\mu})(P_{\nu}+e_0 A_{\nu}) %
         + m^2 e^{2a\phi} \,\Bigr]\,\varphi = 0\,  ,
\label{eq:004}
\end{equation} 
where $\varphi$ possesses information on the ``dynamics'' of the collective
behavior. So the action which yields this wave equation is
\begin{eqnarray}
 S_m\!&=&\!\int\!\!d^4
x\,\sqrt{-g}\:\Bigl[-\,\varphi^{*}e^{-a\phi}g^{\mu\nu} %
               (P_{\mu} + e_0 A_{\mu})(P_{\nu} + e_0 A_{\nu})\varphi %
        - m^2 e^{a\phi} \varphi^{*}\varphi \Bigl]\, ,
\label{eq:005}
\end{eqnarray}
where we regard $\varphi$ as a ``field'' from now on. Therefore the total
action including long-range interactions is
\begin{eqnarray}
 S &=& \int\!\! d^4
x\,\frac{\sqrt{-g}}{16\pi}\:\Bigl[\,R-2 \nabla_{\mu}\phi%
           \nabla^{\mu}\phi - e^{-2a\phi} F_{\mu\nu}F^{\mu\nu}\Bigl] %
          \:+ \;S_m\, ,
\label{eq:006}
\end{eqnarray}
where $R$ is the scalar curvature and
$F_{\mu\nu}=\partial_{\mu}A_{\nu}-\partial_{\nu}A_{\mu}$. The Newton
constant is set to unity.

This action leads to the following field equations:
\begin{eqnarray}
\label{eq:007}
 &&  \nabla^2 \phi + \frac{a}{2}\, e^{-2a\phi} F^2  %
   +\,4\pi a \Bigl[\,e^{-a\phi} \,\varphi^{*} (P+e_0 A)^2 \varphi %
  - e^{a\phi} m^2 \,\varphi^{*} \varphi\, \Bigr] = 0\, ,\\
 &&R_{\mu\nu} - \frac{1}{2}\,g_{\mu\nu}R %
   = 2\,\Bigl[\,\nabla_{\mu}\phi\nabla_{\nu}\phi%
       -\frac{1}{2}\,g_{\mu\nu}(\nabla\phi)^2 \Bigr]  %
+ \:e^{-2a\phi}\Bigl[\,2\,F_{\mu\nu}^2 - \frac{1}{2}\,g_{\mu\nu}F^2 \Bigr]
\nn
 && \qquad\qquad\qquad\qquad
  + 16\pi\biggl\{e^{-a\phi} Re\Bigl[\varphi^{*}(P_{\mu}+e_0 A_{\mu}) %
            (P_{\nu} + e_0 A_{\nu})\,\varphi \nn
 && \label{eq:008} \qquad\qquad\qquad\qquad\qquad\qquad
 - \frac{1}{2}\,g_{\mu\nu}\varphi^{*}(P + e_0 A)^2 \varphi\Bigr] %
- \frac{1}{2}\,g_{\mu\nu} e^{a\phi} m^2 \varphi^{*} \varphi\, \biggr\}\, ,
\\
&&  \nabla_{\mu}\Bigl[\,e^{-2a\phi} F^{\mu\nu} \Bigr] = 8\pi
e_0\,e^{-a\phi}
             \varphi^{*}g^{\nu\lambda}(P_{\lambda} + e_0
A_{\lambda})\,\varphi\, ,
\label{eq:009}
\end{eqnarray}
where $F^2=F_{\mu\nu}F^{\mu\nu}$ and
$F^2_{\mu\nu}=F_{\mu\lambda}F_{\nu}{}^{\lambda}$. $R_{\mu\nu}$ is the Ricci
tensor.

Now, in order to consider only the lowest order in a typical velocity $v$
of  ``extreme black holes'', we assume the following ans\"atze:
\footnote{These ans\"atse are the same as those in
\cite{GiRu,FeEa,TrFe,Shi1,Shi2}.}
\begin{eqnarray}
\label{eq:010}
 ds^2 &=& - U^{-2}\bigl(dt + B_i dx^i \bigr)^2 + U^2 d \Bx^2 \:\: , \\
\label{eq:011}
 U(\Bx) &=& V(\Bx)^{\frac{1}{1+a^2}} \:\: , \\
\label{eq:012}
 e^{-2a\phi} &=& V^{\frac{2 a^2}{1+a^2}} \:\: , \\
\label{eq:013}
 A_0(\Bx) &=& \frac{1}{\sqrt{1+a^2}}\Bigl(1 -\frac{1}{V}\Bigr)  \:\: , \\
\label{eq:014}
 A_i(\Bx) \sim B_i(\Bx)  &=&  O(v)\, ,
\end{eqnarray}
where $i, j,\ldots$ denotes the spatial indices.

If there is no matter source and the vacuum is static
($A_i = B_i = 0$), the ans\"atze together with the mass-charge relation
\begin{equation}
 \frac{e_0}{m} = \sqrt{1 + a^2}\, ,
\label{eq:015}
\end{equation}
 reduce the field equations to
\begin{equation}
 \partial^2 V = 0\, ,
\label{eq:016}
\end{equation}
which implies that the vacuum solution represents an arbitrary number of 
``extreme black holes''\cite{Papa,Madj,Myer,Shir}. One can find that the 
relation between mass $m$ and electric charge $e_0$ of a particle 
corresponds to ``extreme black holes''.
Thus we take this relation~(\ref{eq:015}) hereafter.

Since we consider the low energy limit, $ - \,P_0 - m = E - m \ll m$,
we find
\begin{eqnarray}
 P_0 + e_0 A_0 &=& P_0 + \frac{e_0}{\sqrt{1+a^2}}\,\Bigl(1-\frac{1}{V}\Bigr)
                           = P_0 + m\Bigl(1-\frac{1}{V}\Bigr)  \nn
             &\approx& - \,m \,\frac{1}{V}\, ,
\label{eq:017} \\
 P_i + e_0 A_i - B_i(P_0 + e_0 A_0) &\approx& P_i + %
       e_0\,\Bigl(A_i + \frac{1}{\sqrt{1+a^2}}\,\frac{1}{V}\,B_i\Bigr)  \nn
             &\equiv& P_i + e_0 \hat{A}_i\, ,  \label{eq:018}
\end{eqnarray}
where
\begin{equation}
\hat{A}_i \equiv A_i + \frac{1}{\sqrt{1+a^2}}\,\frac{1}{V}\,B_i \, ,
\label{eq:019}
\end{equation}
and we define
\begin{eqnarray}
\hat{F}_{ij}  &\equiv& \partial_{i}\hat{A}_j - \partial_{j}\hat{A}_i \nn
             &=& \bar{F}_{ij} +
\frac{1}{\sqrt{1+a^2}}\,\frac{1}{V}\,G_{ij}\, ,
\label{eq:020}
\end{eqnarray}
where
\begin{eqnarray}
\label{eq:021}
\bar{F}_{ij}  &\equiv& F_{ij} + B_i F_{j0} - B_j F_{i0}\, , \\
\label{eq:022}
G_{ij}          &\equiv& \partial_{i}B_j - \partial_{j}B_i\, .
\end{eqnarray}

Taking the low energy or non-relativistic limit %
$- P_0 - m = E - m \ll m$, $|P_i + e\hat{A}_i|^2 \approx m^2 v^2 \ll m^2$,
we simplify the field equations (\ref{eq:007}), (\ref{eq:008})
and (\ref{eq:009}) explicitly. From the dilaton field equation
(\ref{eq:007}), the time-time component of the gravitational field equation
(\ref{eq:008}), and the temporal component of the electromagnetic field
equation (\ref{eq:009}), in the lowest order, we obtain
\begin{equation}
\partial^2 V + 8\pi (1 + a^2) m^2 U^3 |\varphi|^2 = 0\, .
\label{eq:023}
\end{equation}

From the time-space component of the gravitational field equation
(\ref{eq:008}), we get
\begin{eqnarray}
\partial_{\ell}\Bigl[V^{\frac{2(a^2-1)}{1+a^2}}\,\frac{1}{V^2}\,G_{\ell i}
\Bigr]&=&
-4\,\sqrt{\frac{1}{1+a^2}}\,\partial_{\ell}\Bigl[V^{\frac{2(a^2-1)}{1+a^2}}\,%
                                \frac{1}{V}\,\bar{F}_{\ell i}\Bigr]  %
    +\,4\,\sqrt{\frac{1}{1+a^2}}\,\frac{1}{V}\,\partial_{\ell}\Bigl[%
           V^{\frac{2(a^2-1)}{1+a^2}}\,\bar{F}_{\ell i}\Bigr]  \nn
&& - \:32\pi\,\frac{m}{V}\;e^{-a\phi}\,\varphi^{*}(P_{i} + e_0
A_{i})\,\varphi\, ,
\label{eq:024}
\end{eqnarray}
On the other hand, the spatial component of the electromagnetic field
equation (\ref{eq:009}) reads
\begin{equation}
\partial_{\ell}\Bigl[V^{\frac{2(a^2-1)}{1+a^2}}\,\bar{F}_{\ell i}\Bigr] %
= 8\pi e_0\,e^{-a\phi}\,\varphi^{*}(P_{i} + e_0 A_{i})\,\varphi\, .
\label{eq:025}
\end{equation}
Here we define an antisymmetric tensor field $H_{ij}$ as
\begin{equation}
\label{eq:026}
H_{ij} \equiv 4\,\sqrt{\frac{1}{1+a^2}}\bar{F}_{ij} + \frac{1}{V}\,G_{ij}.
\end{equation} 
{}From Eqs.~(\ref{eq:024}) and (\ref{eq:025}), the equation for the
antisymmetric tensor field is found to be
\begin{equation}
\partial_{\ell}\Bigl[V^{\frac{2(a^2-1)}{1+a^2}}\,\frac{1}{V}\,H_{\ell i}\Bigr] = 0.
\label{eq:027}
\end{equation} 
Note that from Eqs.~(\ref{eq:020}) and (\ref{eq:026}), we can obtain the
following relation:
\begin{equation}
\frac{1}{4}\,\frac{1}{V^2}\,G^2 - \bar{F}^2 = %
\frac{1+a^2}{3-a^2}\,\Bigl(\hat{F}^2 - \frac{1}{4}\,H^2\Bigr)\, ,
\label{eq:028}
\end{equation} 
where we assumed $a^2\Neq 3$.

Now we consider the effective lagrangian up to the lowest order. Taking the
above estimations into the total action (\ref{eq:006}), we find:
\begin{equation}
S=\int d^4x {\cal L}\, ,
\end{equation}
where the effective lagrangian density is
\begin{eqnarray}
{\cal L} &=& \sqrt{-g}\,\biggl\{\frac{1}{16\pi}\:\Bigl[\,R  %
                          - 2\,(\nabla\phi)^2 - e^{-2a\phi}\,F^2\Bigl] \nn
 &&\qquad\qquad\quad
 + \:\Bigl[- \,\varphi^{*} e^{-a\phi} g^{\mu\nu}(P_{\mu} + e_0
A_{\mu})(P_{\nu} %
   + e_0 A_{\nu})\,\varphi %
   - m^2 e^{a\phi} \varphi^{*}\varphi \Bigl] \biggr\} \nn
 &\approx& 
  \frac{1}{16\pi}\,V^{\frac{2(a^2-1)}{1+a^2}}\,%
    \biggl[\frac{1}{4}\,\frac{1}{V^2}\,G^2 - \bar{F}^2 \biggr] \nn
 && \:
 + \:\Bigl[V\,U^3\varphi^{*}\bigl(P_{0} + e_0 A_{0}\bigr)^2\varphi %
             - \frac{1}{V}m^2 U^3 \varphi^{*}\varphi %
 - \frac{1}{V}\,V^{\frac{2(a^2-1)}{1+a^2}}\,U^3 \varphi^{*}%
                     \bigl(P_{i} + e_0 \hat{A}_{i}\bigr)^2\varphi \Bigr] \nn
 &=& 
 \frac{1}{16\pi}\,V^{\frac{2(a^2-1)}{1+a^2}}\,%
 \biggl[\frac{1+a^2}{3-a^2}\,\Bigl(\hat{F}^2 - \frac{1}{4}\,H^2\Bigr)\biggr]
\nn
\label{eq:030} && \:
 + \:\Bigl[VU^3\varphi^{*}\bigl(P_{0} + e_0 A_{0}\bigr)^2\varphi %
             - \frac{1}{V}m^2 U^3 \varphi^{*}\varphi %
 - \frac{1}{V}\,V^{\frac{2(a^2-1)}{1+a^2}}\,U^3 \varphi^{*}%
                     \bigl(P_{i} + e_0 \hat{A}_{i}\bigr)^2\varphi \Bigr]\, .
\end{eqnarray}
In the first term of this effective lagrangian, $H_{ij}$ can be regarded 
as an independent field, for $H_{ij}$ does not couple to the scalar field
$\varphi$. Because we consider only the interactions among black holes,
we can set $H_{ij}\equiv 0$.

Then Eq.~(\ref{eq:024}), which comes from the time-space component of the
gravitational field equation (\ref{eq:008}), and Eq.~(\ref{eq:025}), from
the spatial component of the electromagnetic field equation (\ref{eq:009}),
can be read as
\begin{equation}
-\,\frac{1+a^2}{3-a^2}\,\partial_{\ell}\Bigl[V^{\frac{2(a^2-1)}{1+a^2}}\,\hat{F}_%
{\ell i}\Bigr] = 8\pi
e_0\,e^{-a\phi}\,\varphi^{*}(P_{i}+e_0\hat{A}_{i})\,\varphi\, ,
\label{eq:031}
\end{equation}
where we assumed $a^2\Neq 3$. For $a^2=3$, the scalar field does not couple
to the force field. For a while, we take $a^2\Neq 3$.

To proceed further, we introduce a non-relativistic field $\psi$:
\begin{equation}
\psi \equiv \sqrt{2m}\,U^{3/2}\,\varphi\, ,
\label{eq:032}
\end{equation}
where for the spatial measure $(g^{(3)})^{1/4} = U^{3/2}$, we obtain a
correct measure for a usual spatial volume. 

Finally, using the approximation
\begin{eqnarray}
\left(P_0+e_0A_0\right)^2&=&
\left(-P_0-m+\frac{m}{V}\right)^2 \nn
&\approx&\frac{m^2}{V^2}+2\frac{m}{V}\left(-P_0-m\right)\, ,
\end{eqnarray}
together, we get the effective lagrangian density in the
low energy limit:
\begin{eqnarray}
{\cal L} &=& 
   \psi^{*}\,\bigl(-\,P_{0} - m\bigr)\psi - %
            \frac{1}{2m\,V^{(3-a^2)/(1+a^2)}}\,\psi^{*}\,\bigl(\BP %
                        + e_0 \hat{\BA}\bigr)^2\psi \nn
&& \qquad + \frac{1}{16\pi}\,\frac{1+a^2}{3-a^2}\,%
\frac{1}{V^{2(1-a^2)/(1+a^2)}}\,\hat{F}^2\, \qquad (a^2\Neq 3) ,
\label{eq:034}
\end{eqnarray}  
where $V$ satisfies the following equation:
\begin{equation}
\partial^2 V + 4\pi\,(1 + a^2)\,m\,|\psi|^2 = 0\, .
\label{eq:035}
\end{equation} 
As a check, varying this effective lagrangian (\ref{eq:034})
with respect to 
$\hat{\BA}$, we can derive again the field equation (\ref{eq:031}) 
in the low energy approximation:
\begin{eqnarray}
-\,\frac{1+a^2}{3-a^2}\,\partial_{\ell}\Bigl[V^{\frac{2(a^2-1)}{1+a^2}}\,\hat{F}_%
{\ell i}\Bigr]&=&4\pi
\frac{e_0}{m}\,V^{\frac{a^2-3}{1+a^2}}\,\psi^{*}(P_{i}+e_0
\hat{A}_{i})\,\psi \nn  
&=&8\pi e_0\,e^{-a\phi}\,\varphi^{*}(P_{i}+e_0
\hat{A}_{i})\,\varphi\, \qquad (a^2\Neq 3)\, .
\label{eq:036}
\end{eqnarray}

For $a^2=3$, since the scalar field does not couple to the vector potential,
the effective lagrangian density at the lowest order is
\begin{equation}
{\cal L}= 
   \psi^{*}\,\bigl(-\,P_{0} - m\bigr)\psi - %
\frac{1}{2m}\,\psi^{*}\,\BP^2\psi\, \qquad (a^2=3)\, ,
\end{equation}  
which seems to describe a free field in the low energy limit.

\section{Gas of ``Extreme Black Holes'' at Finite Temperature}
\label{sec:3}

In this section, we apply the effective theory to the study of the thermal
system of ``extreme black holes''. First we rewrite the effective lagrangian
(\ref{eq:034}) derived in the preceding section as follows:
\footnote{We assume the case of $a^2\Neq 3$, unless
particularly indicated.}
\begin{equation}
{\cal L} =
\psi^{*}\,i\frac{\partial}{\partial t}\,\psi -
 \frac{1}{2\tilde{m}}\,\psi^{*}\, 
\bigl(\BP+\tilde{e}\hat{\BA}\bigr)^2 \psi+\frac{1}{16\pi}\,\hat{F}^2\, ,
\label{eq:038}
\end{equation}
where we have used the notations:
\begin{eqnarray}
\label{eq:039}
\tilde{m} &\equiv& m\, V^{\frac{3-a^2}{1+a^2}}\, , \\
  \tilde{e}^2  &\equiv& \frac{3-a^2}{1+a^2}\,V^{\frac{2(1-a^2)}{1+a^2}}
e_0^2 \nn
\label{eq:040} 
              &=& (3-a^2) V^{\frac{2(1-a^2)}{1+a^2}}\, m^2\, .
\end{eqnarray}
Here we have used the relation between mass $m$ and electric charge $e_0$ is
again
\begin{equation}
\frac{e_0}{m} = \sqrt{1 + a^2}\, .
\label{eq:041}
\end{equation}

Next we consider the field theory at finite temperature\cite{FeWa}.
In this approach, we assume that $V$ is nearly constant and ignore the
problem of operator ordering. 
We define the propagator of the vector field $\hat{\BA}$:
\begin{equation}
\nu_{ij}(\Bq)=\nu(\Bq)\Bigl(\delta_{ij}-\frac{q_i q_j}{q^2}\Bigr)\, .
\label{eq:042}
\end{equation}
In the lowest order, the self-energy of the scalar field, here assumed
to follow Bose-Einstein statistics, is
\footnote{There is no tadopole contribution, because of the derivative
coupling.}
\begin{eqnarray}
 \Sigma(\Bk) &=& -\frac{1}{\beta{\cal V}}\frac{1}{4\tilde{m}^2}
\sum_{q,\ell}\frac{\nu(\Bk-\Bq)}{i\omega_\ell - \epsilon'_q} 
        \Biggl[(\Bk+\Bq)^2 -\frac{\{(\Bk+\Bq)\cdot %
               (\Bk-\Bq)\}^2}{(\Bk-\Bq)^2}\Biggr]\, ,
\label{eq:043} 
\end{eqnarray}
where ${\cal V}$ is the volume of the system, 
$\beta = 1/T$ ($T$ is the temperature of the system) and 
$\omega_{\ell}\:=\:2\ell\pi/\beta$.
$\epsilon'_k$ is considered as
\begin{equation}
\epsilon'_k =  \frac{1}{2\tilde{m}}\Bk^2 + \Sigma(\Bk) -\mu\, ,
\label{eq:044}
\end{equation}
where $\mu$ denotes the chemical potential for the scalar field. 
In the lowest order in interactions, $\nu(\Bq) = \tilde{e}^2/\Bq^2$.
Recalling that we have assumed that the variation of the background field,
or, that of $V$ is small.

We then transform the sum over $\Bq$ into an integral representation:
\begin{equation}
\frac{1}{{\cal V}}\,\sum_{\Bq}\qquad \Longrightarrow %
                       \qquad     \int\!\frac{d^3\Bq}{(2\pi)^3}\, ,
\label{eq:045}
\end{equation} 
and the sum over $\ell$ in this case yields
\begin{equation}
\sum_{\ell}\frac{1}{i\omega_{\ell}-x} = - \frac{\beta}{e^{\beta x}-1}.
\label{eq:046}
\end{equation}

Using these, we can rewrite the self-energy of the scalar field
(\ref{eq:043}) as
\begin{eqnarray}
 \Sigma(\Bk) &=& \frac{\tilde{e}^2}{4\tilde{m}^2}\frac{1}{(2\pi)^3}
\int\! d^3\Bq \,f_q \frac{1}{(\Bk-\Bq)^2}        
 \Biggl[(\Bk+\Bq)^2 - \frac{\{(\Bk+\Bq)\cdot 
                  (\Bk-\Bq)\}^2}{(\Bk-\Bq)^2}\Biggr]   \nn
 &=&\frac{\tilde{e}^2}{\tilde{m}^2}\frac{1}{4\pi^2}\int_0^\infty\!\!   
          dq \,q^2 f_q \Biggl[-1-\frac{k^2+q^2}{2kq} 
                         \ln \left|\frac{k-q}{k+q}\right|\Biggr]\,  ,
\label{eq:047} 
\end{eqnarray}
where $f_q$ is a distribution function of the scalar field:
\begin{equation}
 f_q \equiv \frac{1}{e^{\beta \epsilon'_q} - 1}\, .
\label{eq:048} 
\end{equation}
For a small $k$, corresponding to the low energy, we find:
\begin{equation}
 -1-\frac{k^2+q^2}{2kq}\ln \left|\frac{k-q}{k+q}\right| \approx
\;\frac{4k^2}{3q^2}\, .
\label{eq:049} 
\end{equation}
Since there is no constant term in terms of $k$, 
we can write the self-energy approximately as
\begin{equation}
 \Sigma(\Bk) \:\approx \:\frac{1}{2}\,B k^2\, ,
\label{eq:050} 
\end{equation} 
where $B$ is a constant.

From now on we consider the high-temperature approximation.
\footnote{The particles are still non-relativistic and obey the
Maxwell-Boltzmann distribution.} 
Then the distribution function of the scalar field is reduced to
\begin{equation}
f_q \approx \;e^{-\beta \epsilon'_q}\, .
\label{eq:051} 
\end{equation}
Therefore the self-consistent equation is expressed as
\begin{eqnarray}
 \Sigma(\Bk) &\approx& \;\frac{1}{2}\,B k^2 \nn
            &\approx& 
\frac{\tilde{e}^2}{4 \pi^2\tilde{m}^2}\int_0^\infty\!dq\,q^2
f_q\,\frac{4k^2}{3q^2} \nn
  &=&
\frac{1}{2}\:\frac{4\,\tilde{e}^2\,e^{\beta\mu}}{3(2\pi)^{3/2}\tilde{m}^2}\, 
 \Bigl[\beta\bigl(\frac{1}{\tilde{m}} + B\bigr)\Bigr]^{-1/2}\,k^2\, .
 \label{eq:052} 
\end{eqnarray} 
Equivalently, we get an equation for the coefficient $B$:
\begin{equation}
 B = \frac{4\,\tilde{e}^2\,e^{\beta\mu}}{3(2\pi)^{3/2}\tilde{m}^2}\, 
\Bigl[\beta\bigl(\frac{1}{\tilde{m}} + B\bigr)\Bigr]^{-1/2}\, .
\label{eq:053} 
\end{equation} 
The solution of Eq.~(\ref{eq:053}) is 
\begin{equation}
B = \frac{4\,\tilde{e}^2\,e^{\beta\mu}}{3(2\pi)^{3/2}\tilde{m}^2}\,
\sqrt{\frac{\tilde{m}}{\beta}}\Bigl(\cos \frac{\theta}{3}+   
               \frac{1}{\sqrt{3}}\sin\frac{\theta}{3}\Bigr)^{-1}\, ,
\label{eq:054} 
\end{equation} 
where
\begin{equation}
 \theta \:= \:\arcsin %
\frac{2\sqrt{3}\,\tilde{e}^2 e^{\beta\mu}}{(2\pi)^{3/2}
\sqrt{\beta\tilde{m}}}\, .
\label{eq:055} 
\end{equation}

On the other hand, the particle density $\rho$ is expressed as
\begin{eqnarray}
\rho &=& \frac{1}{(2\pi)^3}\int\! d^3\Bq \,f_q \nn
  &=&     \frac{e^{\beta\mu}}{(2\pi)^{3/2}}\,
\Bigl[\beta\bigl(\frac{1}{\tilde{m}}+B\bigr)\Bigr]^{-3/2}\, .
\label{eq:056} 
\end{eqnarray}
In our model the particle density $\rho$ can also be written by $|\psi|^2$,
therefore
\begin{equation}
 |\psi|^2= \rho = %
    e^{\beta\mu}\Bigl(\frac{\tilde{m}}{2\pi\beta}\Bigr)^{3/2}\,\Bigl[%
    \cos\frac{\theta}{3} + 
\frac{1}{\sqrt{3}}\sin\frac{\theta}{3}\Bigr]^{-3}\, .
\label{eq:057}
\end{equation}
Now we have to solve the following equation with
$|\psi|^2$ given by Eq.~(\ref{eq:057})
\begin{equation}
\partial^2 V + 4\pi\,(1 + a^2)\,m\,|\psi|^2 = 0\, .
\label{eq:058}
\end{equation}

To explicitly solve Eq.~(\ref{eq:058}), let us assume spherical 
symmetry and define the following parameters and the radial coordinate:
\begin{eqnarray}
\label{eq:059}
\rho_0 &\equiv& e^{\beta\mu} \Bigl(\frac{m}{2\pi\beta}\Bigr)^{3/2}\, ,\\
\label{eq:060}
\delta &\equiv& 2\sqrt{3}\,G \,\rho_0 \beta\, , \\
\label{eq:061}
\tilde{r} &\equiv& \sqrt{G \,m \,\rho_0}\,r\, ,
\end{eqnarray} 
where $G$ is the Newton constant, being set to unity until now.
Incidentally, $n_0$ is the particle density of the ideal gas of
particles with mass $m$. Then Eq.~(\ref{eq:058}) for $V$ is
\begin{eqnarray}
\frac{1}{\tilde{r}^2}\Bigl(\tilde{r}^2 V(\tilde{r})^{'}\Bigr)'  %
    &=& -4\pi\,(1 + a^2)V(\tilde{r})^{\frac{3(3-a^2)}{2(1+a^2)}} %
               \Bigl(\, \cos\frac{\theta}{3} %
 + \frac{1}{\sqrt{3}}\sin\frac{\theta}{3}\,\Bigr)^{-3}\, ,
\label{eq:062}
\end{eqnarray}
where the prime denotes derivative with respect to $\tilde{r}$, and
\begin{equation}
 \theta(\tilde{r}) \:= \:\arcsin\Bigl[ %
(3-a^2) V(\tilde{r})^{\frac{(1-3a^2)}{2(1+a^2)}}\, \delta\,\Bigr]\, .
\label{eq:063} 
\end{equation}

\begin{figure}
\centering
(a)\\
\epsfbox{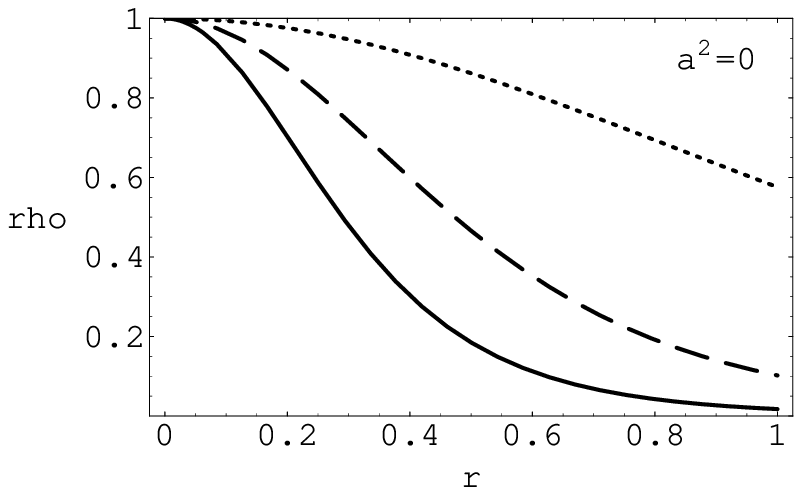}
~\\
(b)\\
\epsfbox{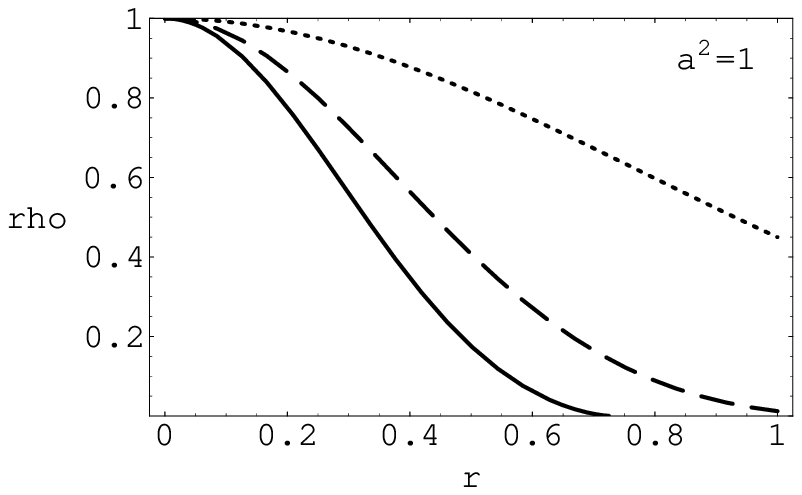}
~\\
(c)\\
\epsfbox{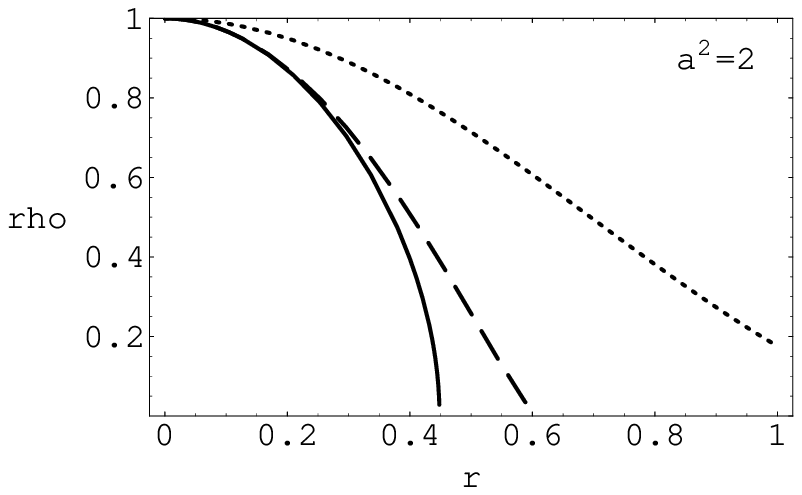}
\caption{%
The density distribution of the isothermal sphere of ``extreme black
holes''  for different values of the coupling constant of the dilaton
field; (a) $a^2=0$,  (b) $a^2=1$ and (c) $a^2=2$. The solid line denotes
$\delta=0$ (the high temperature limit), the broken line $\delta=1$ and
the dotted line $\delta=10$.}
\label{fig:ebh}
\end{figure}

We can numerically solve Eq.~(\ref{eq:062}) with Eq.~(\ref{eq:063})
to obtain $V(\tilde{r})$ and the particle density $\rho=|\psi|^2$ of the
isothermal sphere of ``extreme black holes''.  Fig.~\ref{fig:ebh} shows the
profile of the particle density of the isothermal sphere. The central
density is normalized to unity there. As seen from Fig.~\ref{fig:ebh}, as
$\delta$ is the smaller, that is, the temperature is the higher, the
particles are bound the more tightly. 

We find also that as the coupling
constant of the dilaton field is the larger, the particles are bound the
more tightly. In particular, there is a critical value for the
asymptotic behavior of $n$ at a large distance: for $a^2>1$, the
isothermal sphere at high temperature has a clear ``edge''.

For the case $a^2=3$, the effective lagrangian represents the free
lagrangian, therefore the particles correspond to the free particles.

The physical interpretation on the behavior is given by the fact that the
``magnetic'' force acts attractively between the particles which move in
the opposite direction for $a^2<3$ (see the relative
sign of the coefficient in Eq.~(\ref{eq:036})). The thermal average of
$v^2$ governs the strength of the force; this feature has been suggested in
\cite{ShMa}.

We have also seen that there is the
critical point at $a^2\sim 1$ whether an extent of the ``edge'' exists or
not. This behavior implies the existence of
an effective repulsion in a small range, because similar behavior can be
found in the model of boson stars with a repulsive force\cite{Jetz}.
This critical nature of the interaction seems to correspond to the fact
that the moduli space of two ``extreme black holes'' has a deficit angle
$\pi$ for $a^2=1$\cite{Shi2}. Further study on this subject is expected.

\section{Eliminating the potentials}
\label{sec:4}

We consider the total energy of the system of ``extreme black holes''.
We can eliminate the field stength and the potential apparently in
the expression of the energy using the solution of the field equations
expressed by the scalar field. Assuming localized scalar distributions
in the resulting expression, we can also obtain the energy for the classical
system. For a while, we assume $a^2\Neq 3$.

The total energy of the system is written by
\begin{equation}
H=\int d^3\Bx~{\cal H}\, ,
\end{equation}
with
\begin{equation}
{\cal H} = 
\frac{1}{2m\,V^{(3-a^2)/(1+a^2)}}\,\psi^{*}\,
\bigl(\BP+e_0 \hat{\BA}\bigr)^2\psi
 - \frac{1}{16\pi}\,\frac{1+a^2}{3-a^2}\,
\frac{1}{V^{2(1-a^2)/(1+a^2)}}\,\hat{F}^2\, ,
\end{equation}  
where $e_0=\sqrt{1+a^2}\,m$.

First, we treat the second term of the Hamiltonian density ${\cal H}$. We
use the following definition:
\begin{eqnarray}
\BJ(\Bx) = \frac{\partial {\cal H}}{\partial \BP} &=&
\frac{1}{m\,V^{(3-a^2)/(1+a^2)}}\,\psi^{*}\,\bigl(\BP 
                        + e_0 \hat{\BA}\bigr)\psi \nn
&\equiv& \psi^*(\Bx) \Bv(\Bx) \psi(\Bx)\, .
\label{eq:066}
\end{eqnarray}
Then the field equation~(\ref{eq:036}) can be written as
\begin{equation}
-\,\frac{1+a^2}{3-a^2}\,\partial_{\ell}\Bigl[V^{\frac{2(a^2-1)}{1+a^2}}\,
\hat{F}_{\ell i}\Bigr]=4\pi e_0\,\BJ\, .
\end{equation}
Note that
\begin{equation}
\Bnabla\cdot\BJ=\partial_i \, J^i=0\, .
\end{equation}

As long as $\BJ$ vanishes rapidly at the spatial infinity, the solution is
given by
\begin{eqnarray}
-\frac{1+a^2}{3-a^2}V^{\frac{2(a^2-1)}{1+a^2}}\,
\hat{F}_{\ell i} &=& e_0\,\int d^3\Bx'
\frac{{r'}^{\ell} J^i(\Bx')-{r'}^{i} J^{\ell}(\Bx')}{|\Br'|^3}\, ,
\end{eqnarray}
where $\Br'=\Bx-\Bx'$.
Substituting this solution, we obtain
\begin{eqnarray}
&& \int d^3\Bx \frac{1}{16\pi}\,\frac{1+a^2}{3-a^2}\,
\frac{1}{V^{2(1-a^2)/(1+a^2)}}\,\hat{F}^2 \nn
&=&
\frac{1}{8\pi} \frac{3-a^2}{1+a^2} e_0^2\,\int d^3\Bx \,
V^{\frac{2(1-a^2)}{1+a^2}}(\Bx) \nn 
&& \times\int d^3\Bx' \int d^3\Bx''
\frac{\Br' \cdot \Br'' ~\BJ(\Bx')\cdot\BJ(\Bx'')-
\Br' \cdot\BJ(\Bx'')~\Br'' \cdot\BJ(\Bx')}
{|\Br'|^3|\Br''|^3}\, ,
\label{eq:070}
\end{eqnarray}
where $\Br''=\Bx-\Bx''$.

To garantee the absence of the influence of the spatial infinity 
explicitly, we use the following identity which holds if $\BJ$ vanishes
rapidly at the spatial infinity:
\begin{equation}
 \int d^3\Bx'~
\frac{\Br'\cdot\BJ(\Bx')}
{|\Br'|^3} =0\, ,
\end{equation}  
and modify the equation~(\ref{eq:070}) into
\begin{eqnarray}
&& \int d^3\Bx~\frac{1}{16\pi}\,\frac{1+a^2}{3-a^2}\,
\frac{1}{V^{2(1-a^2)/(1+a^2)}}\,\hat{F}^2 \nn
&=&
\frac{1}{8\pi} \frac{3-a^2}{1+a^2} e_0^2\,\int d^3\Bx \,
V^{\frac{2(1-a^2)}{1+a^2}}(\Bx) \nn 
&& \times\int d^3\Bx' \int d^3\Bx''~
\frac{\Br' \cdot \Br'' ~\BJ(\Bx')\cdot\BJ(\Bx'')-
\Br' \cdot\BJ(\Bx'')~\Br'' \cdot\BJ(\Bx')+
\Br' \cdot\BJ(\Bx')~\Br'' \cdot\BJ(\Bx'')}
{|\Br'|^3|\Br''|^3} \nn
&=&
\frac{1}{8\pi} \frac{3-a^2}{1+a^2} e_0^2\,\int d^3\Bx \,
V^{\frac{2(1-a^2)}{1+a^2}}(\Bx) \nn 
&&\qquad\qquad \times\int d^3\Bx' \int d^3\Bx''~
\frac{{r'}^i {r''}^{j}}{|\Br'|^3|\Br''|^3}\,
(\delta_{k\ell}\delta_{ij}-\delta_{i\ell}\delta_{jk}+
\delta_{ik}\delta_{j\ell})\,J^k(\Bx')J^{\ell}(\Bx'')\, .
\end{eqnarray}  

Next, we consider the first term of the Hamiltonian density ${\cal H}$.
If we define
\begin{equation}
\rho(\Bx) = \psi^*(\Bx)\psi(\Bx)\, ,
\end{equation}
then the equation~(\ref{eq:035}), which is now written as
\begin{equation}
\partial^2 V + 4\pi\,(1 + a^2)\,m\,\rho = 0\, ,
\end{equation}
has the solution:
\begin{equation}
V(\Bx)=1 + (1 + a^2)\,m \int d^3\Bx' \frac{1}{|\Br'|} \rho(\Bx')\, .
\label{eq:075}
\end{equation}

We rearrange the first term of ${\cal H}$ as
\begin{eqnarray}
\frac{1}{2m\,V^{(3-a^2)/(1+a^2)}}\,\psi^{*}\,
\bigl(\BP+e_0 \hat{\BA}\bigr)^2\psi 
&=&\frac{1}{2}\,m\,V^{(3-a^2)/(1+a^2)}\,\psi^{*}\Bv^2\psi \nn
&=&\frac{1}{2}\,m\,\psi^{*}\Bv^2\psi+
\frac{1}{2}m\,\bigl[V^{\frac{3-a^2}{1+a^2}}-1\bigr]\,\psi^{*}\Bv^2\psi\, .
\end{eqnarray}  
Using Eq~(\ref{eq:075}), we find
\begin{eqnarray}
&&\frac{1}{2}\,m\int d^3\Bx~
\bigl[V^{\frac{3-a^2}{1+a^2}}(\Bx)-1\bigr]\,
\psi^{*}(\Bx)\Bv^2(\Bx)\psi(\Bx) \nn
&=&\frac{1}{8\pi}\,m\int d^3\Bx~
\bigl[V^{\frac{3-a^2}{1+a^2}}(\Bx)-1\bigr]
\int d^3\Bx'~
\Bnabla\cdot\frac{\Br'}{|\Br'|^3}\psi^{*}(\Bx')\Bv^2(\Bx')\psi(\Bx') \nn
&=&\frac{3-a^2}{8\pi}\,m^2\int d^3\Bx\,
V^{\frac{2(1-a^2)}{1+a^2}}(\Bx) \nn
&&\qquad\qquad \times\int d^3\Bx'\int d^3\Bx''\,
\frac{\Br'\cdot\Br''}{|\Br'|^3|\Br''|^3}\psi^{*}(\Bx'')\psi(\Bx'')
\psi^{*}(\Bx')\Bv^2(\Bx')\psi(\Bx')\, .
\end{eqnarray}

Consequently, the interaction energy can be written as
\begin{eqnarray}
&&\int d^3\Bx~\Bigl\{
\frac{1}{2}\,m\,\bigl[V^{\frac{3-a^2}{1+a^2}}(\Bx)-1\bigr]\,
\psi^{*}(\Bx)\Bv^2(\Bx)\psi(\Bx)-
\frac{1}{16\pi}\,\frac{1+a^2}{3-a^2}\,
\frac{1}{V^{2(1-a^2)/(1+a^2)}}\,\hat{F}^2\Bigr\} \nn
&=&\frac{3-a^2}{8\pi}~m^2\int d^3\Bx~
V^{\frac{2(1-a^2)}{1+a^2}}(\Bx)
\int d^3\Bx'\int d^3\Bx''~
\frac{1}{|\Br'|^3|\Br''|^3} \nn
&&\times\psi^{*}(\Bx')\psi^*(\Bx'')\Bigl\{
\frac{1}{2}\Br'\cdot\Br''|\Bv(\Bx')-\Bv(\Bx'')|^2-
(\Br'\times\Br'')\cdot(\Bv(\Bx')\times\Bv(\Bx''))
\Bigr\}\psi(\Bx')\psi(\Bx'')\, .
\end{eqnarray}
Because we know that there is no interaction for the $a^2=3$ case,
we see that this expression holds for any value of the dilaton coupling
$a^2$.

For $a^2=1$, the expression for the total enegy can be simplified into
\begin{eqnarray}
H &=&
\int d^3\Bx~
\frac{1}{2}\,m\,\psi^{*}(\Bx)\Bv^2(\Bx)\psi(\Bx)
\nn
&&+\frac{1}{8\pi}\,m^2\int d^3\Bx
\int d^3\Bx'\int d^3\Bx''~
\frac{\Br'\cdot\Br''}{|\Br'|^3|\Br''|^3}
\psi^{*}(\Bx')\psi^*(\Bx'')|\Bv(\Bx')-\Bv(\Bx'')|^2
\psi(\Bx')\psi(\Bx'') \nn
&=&\int d^3\Bx~
\frac{1}{2}\,m\,\psi^{*}(\Bx)\Bv^2(\Bx)\psi(\Bx) \nn
&&+\frac{1}{2}~m^2\int d^3\Bx\int d^3\Bx'~
\psi^{*}(\Bx)\psi^*(\Bx')~\frac{|\Bv(\Bx)-\Bv(\Bx')|^2}{|\Bx-\Bx'|}~
\psi(\Bx)\psi(\Bx')\, .
\end{eqnarray}

Now we consider the classical point-particle limit,
\begin{eqnarray}
\BJ(\Bx)&=& \psi^*(\Bx)\Bv(\Bx)\psi(\Bx) = \sum_a \mu_a \Bv_a~
\delta^3(\Br_a)\, , \\
\rho(\Bx)&=& \psi^*(\Bx)\psi(\Bx) = \sum_a \mu_a\,\delta^3(\Br_a)\, ,
\end{eqnarray}
and
\begin{equation}
\psi^*(\Bx)\Bv^2(\Bx)\psi(\Bx) = \sum_a \mu_a \Bv_a^2~
\delta^3(\Br_a)\, ,
\end{equation}
where $\mu_a$ and $\Bv_a$ are the constant which represents the ratio of the
mass and the velocity of the
$a$-th ``extreme black hole'' located at $\Bx_a$,
respectively.  We use the notation $\Br_a=\Bx-\Bx_a$.
Then one can find that the energy of the classical system takes the form:
\begin{eqnarray}
H&=&\sum_a \frac{1}{2}\,m_a\,\Bv^2_a\nn
&&+\frac{3-a^2}{8\pi}\int d^3\Bx~
V^{\frac{2(1-a^2)}{1+a^2}}(\Bx)
\sum_{a b}~
\frac{m_am_b}{|\Br_a|^3|\Br_b|^3} \nn
&&\qquad \times\Bigl\{
\frac{1}{2}\Br_a\cdot\Br_b|\Bv_a-\Bv_b|^2-
(\Br_a\times\Br_b)\cdot(\Bv_a\times\Bv_b)
\Bigr\}\, ,
\label{eq:083}
\end{eqnarray}
with
\begin{equation}
V(\Bx)=1 + (1 + a^2)\sum_{c}~ \frac{m_c}{|\Br_c|}\, ,
\end{equation}
where the individual mass is $m_a\equiv m\mu_a$.
For $a^2=1$, the energy has a simple form:
\begin{equation}
H =
\sum_a~\frac{1}{2}~m_a \Bv^2_a + \frac{1}{2}\sum_{a b}~
m_am_b~\frac{|\Bv_{ab}|^2}{|\Br_{ab}|}\, ,
\end{equation}
where $\Br_{ab}\equiv\Bx_a-\Bx_b$ and $\Bv_{ab}\equiv\Bv_a-\Bv_b$.

Furthermore, we restrict ourselves on the two-body system.
We assume that the velocity of the center of mass ${\bf V}$ vanishes:
\begin{equation}
{\bf V} \equiv  \frac{m_1\Bv_1+m_2\Bv_2}{M} = {\bf 0}\, ,
\end{equation}
where $M=m_1+m_2$. The velocity of the relative motion is defined as
\begin{equation}
{\bf v} \equiv \Bv_1-\Bv_2\, .
\end{equation}
Then Eq.~(\ref{eq:083}) becomes
\begin{equation}
H=\frac{1}{2}\,\mu\,{\bf v}^2
+\frac{3-a^2}{8\pi}\,\mu M\,{\bf v}^2\int d^3\Bx~
V^{\frac{2(1-a^2)}{1+a^2}}(\Bx)\,
\frac{\Br_1\cdot\Br_2}{|\Br_1|^3|\Br_2|^3}\, ,
\label{eq:088}
\end{equation}
with
\begin{equation}
V(\Bx)=1 + (1 + a^2)\biggl[\frac{m_1}{|\Br_1|}+\frac{m_2}{|\Br_2|}
\biggr]\, ,
\end{equation}
where the reduced mass $\mu$ is given by $\mu=m_1m_2/M$.

In general, a naive integration in Eq.~(\ref{eq:088}) diverges.
Therefore we realize that divergent terms proportional to
$\int d^3\Bx~\delta^3(\Bx)/|\Bx|^p\quad (p>0)$ which appear when the
integrand is expanded must be regularized\cite{InPl}. We set them to zero. 
The prescription is equivalent to carrying out the following replacement 
in Eq.~(\ref{eq:088}):
\begin{equation}
V^{\frac{2(1-a^2)}{1+a^2}}(\Bx)\Rightarrow 
\biggl[1+(1+a^2)\frac{m_1}{|\Br_1|}\biggr]^{\frac{2(1-a^2)}{1+a^2}}+
\biggl[1+(1+a^2)\frac{m_2}{|\Br_2|}\biggr]^{\frac{2(1-a^2)}{1+a^2}}-
1\, .
\end{equation}
Then we get\cite{Shi2}
\begin{eqnarray}
H&=&\frac{1}{2}\,\mu \,{\bf v}^2~\biggl\{ 1-\frac{M}{\mu}-
\frac{(3-a^2) M}{r} \nn
&&\qquad\qquad +\frac{M}{m_1}
\biggl[1 + (1 + a^2)\frac{m_1}{r}\biggr]^{\frac{3-a^2}{1+a^2}}+
\frac{M}{m_2}
\biggl[1 + (1 + a^2)\frac{m_2}{r}\biggr]^{\frac{3-a^2}{1+a^2}}
\biggr\}\, ,
\end{eqnarray}
where $r=|\Bx_1-\Bx_2|$.

Before closing this section, we show the results for $(N+1)$ dimensional
case. We assume $a^2\Neq N$ here. The basic action is given in Appendix.
We find that the effective action is
\begin{equation}
{\cal H} = 
\frac{1}{2m\,V^{\frac{N-a^2}{N-2+a^2}}}\,\psi^{*}\,
\bigl(\BP+e_0 \hat{\BA}\bigr)^2\psi
- \frac{1}{16\pi}\,\frac{N-2+a^2}{N-a^2}\,
\frac{1}{V^{\frac{2(1-a^2)}{N-2+a^2}}}\,\hat{F}^2\, ,
\end{equation}  
where
\begin{equation}
e_0^2=\frac{2(N-2+a^2)}{N-1}\, m^2\, .
\end{equation}
Here $V$ satisfies
\begin{equation}
\partial^2 V + 8\pi\,\frac{N-2 + a^2}{N-1}\,m\,|\psi|^2 = 0\, .
\end{equation} 
The derivation of the effective action is shown in Appendix.

Similarly to the previous analysis, we obtain the interaction Hamiltonian:
\begin{eqnarray}
H_{int}&=&\int d^N\Bx~\Bigl\{
\frac{1}{2}m\,\bigl[V^{\frac{N-a^2}{N-2+a^2}}(\Bx)-1\bigr]\,
\psi^{*}(\Bx)\Bv^2(\Bx)\psi(\Bx) \nn
&&\qquad\qquad-\frac{1}{16\pi}\,\frac{N-2+a^2}{N-a^2}\,
\frac{1}{V^{2(1-a^2)/(N-2+a^2)}}\,\hat{F}^2\Bigr\}\nn
&=&\frac{N-a^2}{4(N-1)\pi}\left(\frac{4\pi}{A_{N-1}}\right)^2 m^2\int
d^N\Bx~ V^{\frac{2(1-a^2)}{N-2+a^2}}(\Bx)
\int d^N\Bx'\int d^N\Bx''~
\frac{1}{|\Br'|^N|\Br''|^N} \nn
&&\times\psi^{*}(\Bx')\psi^*(\Bx'')\Bigl\{
\frac{1}{2}\Br'\cdot\Br''|\Bv(\Bx')-\Bv(\Bx'')|^2-
(\Br'\cdot\Bv(\Bx'))(\Br''\cdot\Bv(\Bx'')) \nn
&&\qquad\qquad\qquad\quad +
(\Br'\cdot\Bv(\Bx''))(\Br''\cdot\Bv(\Bx'))\Bigr\}\psi(\Bx')\psi(\Bx'')\, ,
\end{eqnarray}
where
\begin{equation}
\Bv(\Bx) \equiv \frac{1}{m\,V^{(N-a^2)/(N-2+a^2)}}\,
\bigl(\BP+ e_0 \hat{\BA}\bigr)\, ,
\end{equation}
and $A_{N-1}=2\pi^N/\Gamma(N/2)$ is the volume of a unit $N-1$ sphere.

Especially for $a^2=1$, we find
\begin{equation}
H_{int}=\frac{1}{2}\,\frac{4\pi}{A_{N-1}}\,
m^2\int d^N\Bx\int d^N\Bx'~
\psi^{*}(\Bx)\psi^*(\Bx')
\frac{|\Bv(\Bx)-\Bv(\Bx')|^2}{(N-2)|\Bx-\Bx'|^{N-2}}
\psi(\Bx)\psi(\Bx') \, .
\end{equation}

The classical limit of total energy of the system is found to be
\begin{eqnarray}
H&=&\sum_a \frac{1}{2}\,m_a\,\Bv^2_a\nn
&&+\frac{N-a^2}{4(N-1)\pi}\left(\frac{4\pi}{A_{N-1}}\right)^2\int d^N\Bx~
V^{\frac{2(1-a^2)}{N-2+a^2}}(\Bx)
\sum_{a b}~
\frac{m_am_b}{|\Br_a|^N|\Br_b|^N} \nn
&&\qquad \times\Bigl\{
\frac{1}{2}\Br_a\cdot\Br_b|\Bv_a-\Bv_b|^2-
(\Br_a\cdot\Bv_a)(\Br_b\cdot\Bv_b)+(\Br_a\cdot\Bv_b)(\Br_b\cdot\Bv_a)
\Bigr\}\, ,
\end{eqnarray}
with
\begin{equation}
V(\Bx)=1+\frac{2(N-2+a^2)}{(N-1)(N-2)}\frac{4\pi}{A_{N-1}}
\sum_{c}~ \frac{m_c}{|\Br_c|^{N-2}}\, .
\label{eq:099}
\end{equation}
For $a^2=1$, the total energy of the system has a simple form:
\begin{equation}
H =
\sum_a~\frac{1}{2}~m_a \Bv^2_a +
\frac{1}{2}\,\frac{4\pi}{A_{N-1}}\sum_{a b}~
m_am_b~\frac{|\Bv_{ab}|^2}{(N-2)|\Br_{ab}|^{N-2}} \, .
\end{equation}

Finally, we show another expression for the total energy.
That is
\begin{equation}
H =\sum_{a b}v^{ak}v^{b\ell}
(\delta^i_{k}\delta^j_{\ell}+\delta_{k\ell}\delta^{ij}-
\delta^j_{k}\delta^i_{\ell})
\partial_{ai}\partial_{bj}L \, ,
\end{equation}
where
\begin{equation}
L =-\frac{1}{32\pi}\int d^N\Bx~V^{\frac{2(N-1)}{N-2+a^2}}(\Bx) \, ,
\end{equation}
with $V$ given by Eq.~(\ref{eq:099}).

\section{Conclusion}
\label{sec:5}

In this paper, we have derived the effective lagrangian of 
``extreme black holes'' in the low energy limit. At finite
temperature, we have obtained a self-consistent equation and then we have
seen the structure of the isothermal sphere distribution of ``extreme black
holes''.  At high temperature, the gas of ``extreme black holes''  have
been lumped by the velocity-dependent force. 

In future work, we will consider the low-temperature case. 
It is interesting to investigate whether the condensation of ``extreme black
holes'' may take place.  We will also take another posibility for the
statistics of ``extreme black holes'' into consideration. We would like to
study the effective theory by means of the lattice calculation, in which
the strongly coupled system is appropriately treated.

We have also studied the classical point-particle limit for the energy of
the system. Moduli space structure and supersymmetric extensions of the
multiple black hole system have recently been studied by many
authors\cite{GiKa,BRO,GPS,KaMi,Mich,MMS,MiSt,MSS,GP1,GP2,SCQM}. We are
interested in such a direction of study and expect some symmetric structure
to be found in the effective field theory of multi-black holes.

\section*{Acknowledgements}
KS would like to thank Satoru Hirenzaki for reading this manuscript.
He also thank Nahomi Kan and Yoshinori Cho for useful comments.
\newpage

\appendix
\section*{}

In this Appendix, we show the derivation of the effective lagrangian for 
``extreme black holes'' in the $(N+1)$ dimensional spacetime in detail.

We start with the action for the charged scalar field: 
\begin{eqnarray}
\label{eq:A01}
S_m\!&=&\!\int\!\!d^{N+1}
x\,\sqrt{-g}\:\Bigl[-\,\varphi^{*}e^{-\frac{2a}{N-1}\phi}g^{\mu\nu} %
               (P_{\mu} + e_0 A_{\mu})(P_{\nu} + e_0 A_{\nu})\varphi  %
        - m^2 e^{\frac{2a}{N-1}\phi} \varphi^{*}\varphi \Bigl]\, .
\end{eqnarray}

The total action reads
\begin{eqnarray}
\label{eq:A02}
 S &=& \int\!\! d^{N+1}
x\,\frac{\sqrt{-g}}{16\pi}\:\Bigl[\,R-\frac{4}{N-1}\nabla_{\mu}\phi%
           \nabla^{\mu}\phi - e^{-\frac{4a}{N-1}\phi}
F_{\mu\nu}F^{\mu\nu}\,\Bigl] %
          \:+ \;S_m ,
\end{eqnarray}
and leads to the field equations:
\begin{eqnarray}
\label{eq:A03} 
 &&  \nabla^2 \phi + \frac{a}{2}\, e^{-\frac{4a}{N-1}\phi} F^2  %
   +\,4\pi a \Bigl[\,e^{-\frac{2a}{N-1}\phi} \,\varphi^{*} (P + e_0 A)^2
\varphi 
  - e^{\frac{2a}{N-1}\phi} m^2 \,\varphi^{*} \varphi\, \Bigr] = 0\, ,\\
 &&R_{\mu\nu} - \frac{1}{2}\,g_{\mu\nu}R %
   = \frac{4}{N-1}\,\Bigl[\,\nabla_{\mu}\phi\nabla_{\nu}\phi%
       -\frac{1}{2}\,g_{\mu\nu}(\nabla\phi)^2 \Bigr]  %
 + \:e^{-\frac{4a}{N-1}\phi}\Bigl[\,2\,F_{\mu\nu}^2 -
\frac{1}{2}\,g_{\mu\nu}F^2 \Bigr]
\nn
 && \qquad\qquad\qquad\qquad
  + 16\pi\biggl\{e^{-\frac{2a}{N-1}\phi} Re\Bigl[\varphi^{*}(P_{\mu} + e_0
A_{\mu}) (P_{\nu} + e_0 A_{\nu})\,\varphi \nn  
 && \label{eq:A04} \qquad\qquad\qquad\qquad\qquad\qquad
 - \frac{1}{2}\,g_{\mu\nu}\varphi^{*}(P + e_0 A)^2 \varphi\Bigr] %
    - \frac{1}{2}\,g_{\mu\nu} e^{\frac{2a}{N-1}\phi} m^2 \varphi^{*}
\varphi\, \biggr\}\, , \\ &&  \nabla_{\mu}\Bigl[\,e^{-\frac{4a}{N-1}\phi}
F^{\mu\nu}
\Bigr] = 8\pi e_0\,e^{-\frac{2a}{N-1}\phi}
             \varphi^{*}g^{\nu\lambda}(P_{\lambda} + e_0
A_{\lambda})\,\varphi\, .
\label{eq:A05}
\end{eqnarray}

The ans\"atze in the $(N+1)$ dimensional case are now:
\begin{eqnarray}
\label{eq:A06}
 ds^2 &=& - U^{-2}\bigl(dt + B_i dx^i \bigr)^2 + U^{\frac{2}{N-2}} d \Bx^2
\:\: ,
\\
\label{eq:A07}
 U(\Bx) &=& V(\Bx)^{\frac{N-2}{N-2+a^2}} \:\: , \\
\label{eq:A08}
 e^{-\frac{4a}{N-1}\phi} &=& V^{\frac{2 a^2}{N-2+a^2}} \:\: , \\
\label{eq:A09}
 A_0(\Bx) &=& \sqrt{\frac{N-1}{2(N-2+a^2)}}\Bigl(1 -\frac{1}{V}\Bigr) 
\:\: , \\
\label{eq:A10}
 A_i(\Bx) \sim B_i(\Bx)  &=&  O(v)\, .
\end{eqnarray} 

In addition, the following charge-mass ratio is assumed:
\begin{equation}
 \frac{e_0}{m} = \sqrt{\frac{2 (N-2 + a^2)}{N-1}}.
\label{eq:A11}
\end{equation}
This corresponds to that of the ``extreme black holes''.

Now we consider the low energy limit, $ - \,P_0 - m = E - m \ll m$.
Then
\begin{eqnarray}
&& P_0 + e_0 A_0 = 
P_0 + e_0\sqrt{\frac{N-1}{2(N-2+a^2)}}\,\Bigl(1-\frac{1}{V}\Bigr)
= P_0 + m\Bigl(1-\frac{1}{V}\Bigr)\approx - \,m\,\frac{1}{V}\, ,
\label{eq:A12} \\
&& P_i + e_0 A_i - B_i(P_0 + e_0 A_0) \approx
 P_i+e_0\,\Bigl(A_i +
\sqrt{\frac{N-1}{2(N-2+a^2)}}\,\frac{1}{V}\,B_i\Bigr)
\equiv P_i+e_0\hat{A}_i\, ,  \label{eq:A13}
\end{eqnarray}
where
\begin{equation}
\label{eq:A14}
 \hat{A}_i \equiv A_i + \sqrt{\frac{N-1}{2
(N-2+a^2)}}\,\frac{1}{V}\,B_i \, ,
\end{equation}
and we define
\begin{equation}
\hat{F}_{ij}\equiv\partial_{i}\hat{A}_j - \partial_{j}\hat{A}_i 
=\bar{F}_{ij} +
\sqrt{\frac{N-1}{2 (N-2+a^2)}}\,\frac{1}{V}\,G_{ij}\, ,
\label{eq:A15}
\end{equation}
where
\begin{eqnarray}
\label{eq:A16}
\bar{F}_{ij}  &\equiv& F_{ij} + B_i F_{j0} - B_j F_{i0}\, , \\
\label{eq:A17}
G_{ij}          &\equiv& \partial_{i}B_j - \partial_{j}B_i\, .
\end{eqnarray}
Using the ans\"atze and taking the low energy or non-relativistic limit %
$- P_0 - m = E - m \ll m$, $|P_i + e\hat{A}_i|^2 \approx m^2 v^2 \ll m^2$,
we simplify the field equations.

We reduce the dilaton field equation (\ref{eq:A03}), using
Eqs.~(\ref{eq:A06}), (\ref{eq:A13}), (\ref{eq:A14}), and (\ref{eq:A16}), to
\begin{eqnarray}
&&U^{-\frac{2}{N-2}}\partial^2\phi+\frac{a}{2}\,
e^{-\frac{4a}{N-1}\phi}\Bigl[-\,2U^{2}U^{-\frac{2}{N-2}}(F_{0i})^2+
U^{-\frac{4}{N-2}}\bar{F}^2\Bigr]  \nn
 &&  + \,4\pi a \biggl\{\,e^{-\frac{2a}{N-1}\phi}\,\Bigl[-\,U^2
\varphi^{*}(P_0+e_0 A_0)^2 \varphi +
U^{-\frac{2}{N-2}}\varphi^{*}(P_{i} + e_0 \hat{A}_{i})^2 \varphi
\Bigr] \nn && \qquad\qquad\qquad\qquad\qquad\qquad\qquad\qquad
  - e^{\frac{2a}{N-1}\phi} m^2 \,\varphi^{*} \varphi\, \biggr\} = 0\, ,
\label{eq:A18}
\end{eqnarray}
where $\bar{F}^2=\bar{F}_{ij}\bar{F}_{ij}$.
Further using Eqs.~(\ref{eq:A07}), (\ref{eq:A08}), and (\ref{eq:A09}), we
get
\begin{eqnarray}
&&\frac{a(N-1)}{2(N-2+a^2)}\,\frac{1}{V}\partial^2 V
+\,4\pi a\,\Bigl[e^{-\frac{2a}{N-1}\phi}U^2\varphi^{*}(P_0+e_0 A_0)^2
\varphi %
 + e^{\frac{2a}{N-1}\phi}m^2 \,\varphi^{*} \varphi\,\Bigr] U^{\frac{2}{N-2}}
\nn &=& \:
\frac{a}{2}\, e^{-\frac{4a}{N-1}\phi}U^{-\frac{2}{N-2}}\bar{F}^2 %
+4\pi aU^{-\frac{2}{N-2}}\varphi^{*}(P_{i}+e_0\hat{A}_{i})^2\varphi\, .
\label{eq:A19}
\end{eqnarray}
Finally we use (\ref{eq:A12}) and rearrange the equation, and
because the right hand side of Eq.~(\ref{eq:A19}) is $O(v^2)$, we find
that the dilaton equation in the lowest order can be reduced to
\begin{equation}
\partial^2 V +16\pi\,\frac{N-2+a^2}{N-1}\,m^2
U^{\frac{N}{N-2}}|\varphi|^2=0\, .
\label{eq:A20}
\end{equation}

The time-time component of the gravitational field equation (\ref{eq:A04})
can be treated in the same manner. For the first step, we use the metric
ansatz (\ref{eq:A06}) and then get
\begin{eqnarray}
&&
   \frac{N-1}{N-2}\,U^{-\frac{2}{N-2}}\partial_{\ell}\Bigl(\delu{\ell}\Bigr) %
   + \frac{1}{2}\frac{N-1}{N-2}\,U^{-\frac{2}{N-2}}\Bigl(\delu{\ell}\Bigr)^2 %
   - \frac{3}{8}\,U^{-2}U^{-\frac{4}{N-2}}G^2 \nn
&=&
    \frac{4}{N-1}\,\Bigl[-\frac{1}{2}\,U^{-\frac{2}{N-2}}\bigl(\partial_{k}
    \phi\bigr)^2\Bigr]+e^{-\frac{4a}{N-1}\phi}\Bigl[-U^2U^{-\frac{2}{N-2}}
    \bigl(F_{0k}\bigr)^2-\frac{1}{2}U^{-\frac{4}{N-2}}\bar{F}^2\Bigr]
\nn && 
+\,8\pi\biggl\{e^{-\frac{2a}{N-1}\phi}\,\Bigl[-U^2\varphi^{*}%
  (P_0 + e_0 A_0)^2\,\varphi - U^{-\frac{2}{N-2}}\varphi^{*}%
   (P_{i} + e_0 \hat{A}_{i})^2 \varphi\Bigr]
        -e^{\frac{2a}{N-1}\phi} m^2 \varphi^{*} \varphi\, \biggr\}\, ,
\label{eq:A21} 
\end{eqnarray}
where $G^2=G_{ij}G_{ij}$.
Next, we use $V$ and obtain
\begin{eqnarray}
&&\frac{N-1}{N-2+a^2}\frac{1}{V}\,\partial^2 V
+\,8\pi\biggl[e^{-\frac{2a}{N-1}%
   \phi}U^2\varphi^{*}(P_0 + e_0 A_0)^2\,\varphi %
+ e^{\frac{2a}{N-1}\phi} m^2 \varphi^{*} \varphi \biggr]U^{\frac{2}{N-2}}
\nn &=& %
\frac{3}{8}\,U^{-2}U^{-\frac{2}{N-2}}G^2 %
 - \frac{1}{2}e^{-\frac{4a}{N-1}}\bar{F}^2 %
 - 8\pi e^{-\frac{2a}{N-1}\phi}\varphi^{*}(P_{i} 
  + e_0 \hat{A}_{i})^2 \varphi\, . \label{eq:A22}
\end{eqnarray}
Finally we pick up a part of the lowest order. The reduced equation is the
same as Eq.~(\ref{eq:A20}).

The temporal component of the electromagnetic field equation
(\ref{eq:A05}), in the lowest order, can be read as
\begin{eqnarray}
&&
U^{-\frac{2}{N-2}}\partial_{k}\biggl[\,e^{-\frac{4a}{N-1}\phi}
\Bigl(U^2F_{0k}+U^{-\frac{2}{N-2}}B_{\ell}\bar{F}_{\ell
k}\Bigr)\biggr] \nn
 &=&
- 8\pi e_0\,e^{-\frac{2a}{N-1}\phi}\biggl[U^2\varphi^{*}(P_0 + e_0
A_0)\,\varphi+U^{-\frac{2}{N-2}}B_k\,\varphi^{*}(P_{k} + e_0
\hat{A}_{k})\,\varphi\biggr]\, .
 \label{eq:A23}
\end{eqnarray}
This equation is equivalent to
\begin{eqnarray}
&&\sqrt{\frac{N-1}{2(N-2+a^2)}}\,\partial^2 V-8\pi
e_0\,e^{-\frac{2a}{N-1}\phi}%
     U^{\frac{N}{N-2}}U\varphi^{*}(P_0 + e_0 A_0)\,\varphi \nn
 &=&
\partial_{k}\Bigl[\,e^{-\frac{4a}{N-1}\phi}U^{-\frac{2}{N-2}}B_{\ell}%
 \bar{F}_{\ell k}\Bigr]+ 8\pi e_0\,e^{-\frac{2a}{N-1}\phi}B_k\,\varphi^{*}%
  (P_{k} + e_0 \hat{A}_{k})\,\varphi\, . \label{eq:A24}
\end{eqnarray}
The right hand side of this equation is $O(v^2)$. Together
with Eq.~(\ref{eq:A11}), we find that Eq.~(\ref{eq:A24}) reduces to
Eq.~(\ref{eq:A20}) in the lowest order in $v$.

From the time-space component of the gravitational field equation
(\ref{eq:A04}),  we obtain
\begin{eqnarray}
&& -\,\frac{1}{2}\,UU^{-\frac{1}{N-2}}\partial_{\ell}
       \Bigl(U^{-2}U^{-\frac{2}{N-2}}G_{\ell i}\Bigr) \nn
&=& -\,2e^{-\frac{4a}{N-1}\phi}UU^{-\frac{3}{N-2}}F_{0k}\bar{F}_{ik}-
16\pi e^{-\frac{2a}{N-1}\phi}UU^{-\frac{1}{N-2}}\Bigl[\,
        \varphi^{*}\bigl(P_0+ e_0 A_0\bigr) %
        \bigl(P_{i} + e_0 \hat{A}_{i}\bigr)\varphi\Bigr]\, .
\label{eq:A25} 
\end{eqnarray}
This can be reduced to
\begin{eqnarray}
&& \partial_{\ell}\biggl[V^{\frac{2(a^2-1)}{N-2+a^2}}\frac{1}{V^2}%
      \,G_{\ell i}\biggr] \nn
&=&-\,4\sqrt{\frac{N-1}{2(N-2+a^2)}}\,\partial_{k}\biggl[%
   V^{\frac{2(a^2-1)}{N-2+a^2}}\frac{1}{V}\bar{F}_{ki}\biggr] %
    + 4\sqrt{\frac{N-1}{2(N-2+a^2)}}\frac{1}{V}\,\partial_{k}\biggl[%
    V^{\frac{2(a^2-1)}{N-2+a^2}}\bar{F}_{ki}\biggr] \nn
 && %
  +32\pi e^{-\frac{2a}{N-1}\phi}\Bigl[\,\varphi^{*}\bigl(P_0+e_0A_0\bigr)
        \bigl(P_{i} + e_0 \hat{A}_{i}\bigr)\varphi\Bigr] \nn
&\approx&-\,4\sqrt{\frac{N-1}{2(N-2+a^2)}}\,\partial_{k}\biggl[%
   V^{\frac{2(a^2-1)}{N-2+a^2}}\frac{1}{V}\bar{F}_{ki}\biggr] %
    + 4\sqrt{\frac{N-1}{2(N-2+a^2)}}\frac{1}{V}\,\partial_{k}\biggl[%
    V^{\frac{2(a^2-1)}{N-2+a^2}}\bar{F}_{ki}\biggr] \nn
 && %
  -32\pi m\frac{1}{V}
e^{-\frac{2a}{N-1}\phi}\varphi^{*}
        \bigl(P_{i} + e_0 \hat{A}_{i}\bigr)\varphi\, .
\label{eq:A26} 
\end{eqnarray}
On the other hand, the spatial component of the electromagnetic field
equation  (\ref{eq:A05}) reads
\begin{equation}
\partial_{k}\biggl[\,e^{-\frac{4a}{N-1}\phi}U^{-\frac{2}{N-2}} %
 \bar{F}_{ki}\biggr]  %
= 8\pi e_0\,e^{-\frac{2a}{N-1}\phi}\varphi^{*}(P_{k} + e_0
\hat{A}_{k})\varphi\, ,
\label{eq:A27}
\end{equation}
or equivalently,
\begin{eqnarray}
\partial_{k}\biggl[\,V^{\frac{2(a^2-1)}{N-2+a^2}}\bar{F}_{ki}\biggr]  %
= 8\pi e_0\,e^{-\frac{2a}{N-1}\phi}\varphi^{*}(P_{k} + e_0
\hat{A}_{k})\varphi\, .
\label{eq:A28}
\end{eqnarray}
Finally using the mass-charge relation (\ref{eq:A11}), we have
\begin{eqnarray}
\sqrt{\frac{N-1}{2(N-2+a^2)}}\,\partial_{k}
\biggl[\,V^{\frac{2(a^2-1)}{N-2+a^2}}%
             \bar{F}_{ki}\biggr]  %
= 8\pi m\,e^{-\frac{2a}{N-1}\phi}\varphi^{*}(P_{k} + e_0
\hat{A}_{k})\varphi\, .
\label{eq:A29}
\end{eqnarray}


By taking the same low-energy approximation
into the total action (\ref{eq:A02}), we obtain the
effective lagrangian density ${\cal L}$, where
\begin{equation}
S=\int d^{N+1}x \,\,{\cal L} \, .
\end{equation}

 Note that:
\begin{eqnarray}
R &=&
-\frac{2}{N-2}\,U^{-\frac{2}{N-2}}\partial_{\ell}\Bigl(\delu{\ell}\Bigr)
      - \frac{N-1}{N-2}\,U^{-\frac{2}{N-2}}\Bigl(\delu{\ell}\Bigr)^2 +
\frac{1}{4}\,U^{-2}U^{-\frac{4}{N-2}}G^2 \nn
&=& -\frac{2}{N-2}\,U^{-\frac{2}{N-2}}\partial_{\ell}\Bigl(\delu{\ell}\Bigr)
 -\frac{(N-1)(N-2)}{(N-2+a^2)^2}\,U^{-\frac{2}{N-2}}
\Bigl(\frac{\partial_{\ell}V}{V}\Bigr)^2 \nn
&& +
\frac{1}{4}V^{\frac{2(a^2-1)}{N-2+a^2}}U^{-\frac{2}{N-2}}\frac{1}{V^2}G^2\,,
\label{eq:A31}
\end{eqnarray}

\begin{eqnarray}
-\frac{4}{N-1}\nabla_{\mu}\phi\nabla^{\mu}\phi &=&
-\frac{4}{N-1}U^{-\frac{2}{N-2}}(\partial_{\ell}\phi)^2 \nn
        &=&
-\frac{(N-1)a^2}{(N-2+a^2)^2}U^{-\frac{2}{N-2}}
\Bigl(\frac{\partial_{\ell}V}{V}\Bigr)^2\, ,
\label{eq:A32}
\end{eqnarray}

\begin{eqnarray}
-e^{-\frac{4a}{N-1}\phi}F^2&=&V^{\frac{2a^2}{N-2+a^2}}
\bigl[2U^{2}U^{-\frac{2}{N-2}}(F_{0i})^2-
U^{-\frac{4}{N-2}}\bar{F}^2\bigr] \nn
&=&\frac{N-1}{N-2+a^2}U^{-\frac{2}{N-2}}
\Bigl(\frac{\partial_{\ell}V}{V}\Bigr)^2-
V^{\frac{2(a^2-1)}{N-2+a^2}}U^{-\frac{2}{N-2}}\bar{F}^2\, .
\label{eq:A33}
\end{eqnarray} 

Now we find:
\begin{eqnarray}
{\cal L} &=& \sqrt{-g}\,\biggl\{\frac{1}{16\pi}\:\Bigl[\,R  %
                   - \frac{4}{N-1}\,(\nabla\phi)^2 -
e^{-\frac{4a}{N-1}\phi}\,F^2\Bigl] \nn
 &&\qquad\qquad\quad
 + \:\Bigl[- \,\varphi^{*} e^{-\frac{2a}{N-1}\phi} g^{\mu\nu}
(P_{\mu} + e_0 A_{\mu})(P_{\nu}+ e_0 A_{\nu})\,\varphi %
   - m^2 e^{\frac{2a}{N-1}\phi} \varphi^{*}\varphi \Bigl] \biggr\} \nn
 &\approx& 
  \frac{1}{16\pi}\,V^{\frac{2(a^2-1)}{N-2+a^2}}\,%
    \biggl[\frac{1}{4}\,\frac{1}{V^2}\,G^2 - \bar{F}^2 \biggr] \nn
 && \:
 + \:U^{\frac{N}{N-2}}\Bigl[V\,\varphi^{*}\bigl(P_{0} + e_0
A_{0}\bigr)^2\varphi
             - \frac{1}{V}m^2 \varphi^{*}\varphi %
 - \frac{1}{V}\,V^{\frac{2(a^2-1)}{N-2+a^2}}\, \varphi^{*}%
 \bigl(P_{i} + e_0 \hat{A}_{i}\bigr)^2\varphi \Bigr] \nn
 &=& 
 \frac{1}{16\pi}\,V^{\frac{2(a^2-1)}{N-2+a^2}}\,%
 \biggl[\frac{N-2+a^2}{N-a^2}\,\Bigl(\hat{F}^2 -
\frac{1}{4}\,H^2\Bigr)\biggr] \nn
\label{eq:A34} && \:
 + \:U^{\frac{N}{N-2}}\Bigl[V\varphi^{*}\bigl(P_{0} + e_0
A_{0}\bigr)^2\varphi - \frac{1}{V}m^2 \varphi^{*}\varphi %
 - \frac{1}{V}\,V^{\frac{2(a^2-1)}{N-2+a^2}}\, \varphi^{*}
\bigl(P_{i} + e_0 \hat{A}_{i}\bigr)^2\varphi \Bigr]\, .
\end{eqnarray}

Here we have defined an antisymmetric tensor field $H_{ij}$ as
\begin{equation}
H_{ij} \equiv 4\,\sqrt{\frac{N-1}{2 (N-2+a^2)}}\bar{F}_{ij} +
\frac{1}{V}\,G_{ij}\, .
\label{eq:A35}
\end{equation}
$H_{ij}$ does not couple to the scalar field $\varphi$, thus we set $H_{ij}
\equiv 0$.
Then both Eqs.~(\ref{eq:A26}) and (\ref{eq:A29}) can be read as
\begin{equation}
-\,\frac{N-2+a^2}{N-a^2}\,\partial_{\ell}\Bigl[V^{\frac{2(a^2-1)}{N-2+a^2}}\,\hat{F}_%
{\ell i}\Bigr] = 8\pi
e_0\,e^{-\frac{2a}{N-1}\phi}\,\varphi^{*}(P_{i}+e_0\hat{A}_{i})\,\varphi\, .
\label{eq:A36}
\end{equation}

To proceed further, we introduce a non-relativistic field $\psi$:
\begin{equation}
\psi \equiv \sqrt{2m}\,U^{\frac{N}{2(N-2)}}\,\varphi\, ,
\label{eq:A37}
\end{equation}
where since the spatial volume measure $(g^{(N)})^{1/4} =
U^{\frac{N}{2(N-2)}}$, we obtain a correct measure for a usual spatial volume. 

Finally we get the effective lagrangian density in the
low energy limit:
\begin{eqnarray}
{\cal L} &=& 
\psi^{*}\,\bigl(-\,P_{0} - m\bigr)\psi - 
\frac{1}{2m\,V^{(N-a^2)/(N-2+a^2)}}\,\psi^{*}\,\bigl(\BP 
                        + e_0 \hat{\BA}\bigr)^2\psi \nn
&& \qquad + \frac{1}{16\pi}\,\frac{N-2+a^2}{N-a^2}\,
\frac{1}{V^{2(1-a^2)/(N-2+a^2)}}\,\hat{F}^2\, \qquad (a^2\Neq N)\, ,
\label{eq:A38}
\end{eqnarray}  
where $V$ satisfies the following equation:
\begin{equation}
\partial^2 V + 8\pi\,\frac{N-2 + a^2}{N-1}\,m\,|\psi|^2 = 0\, .
\label{eq:A39}
\end{equation} 
Varying this effective lagrangian (\ref{eq:A38}) with respect to 
$\hat{\BA}$, we can derive again the field equation
equivalent to Eq.~(\ref{eq:A36}) in the low-energy approximation.

For $a^2=N$, since the scalar field does not couple to the vector field,
the effective lagrangian density at the lowest order is
\begin{equation}
{\cal L}= 
\psi^{*}\,\bigl(-\,P_{0} - m\bigr)\psi - 
\frac{1}{2m}\,\psi^{*}\,\BP^2\psi \qquad (a^2=N)\, .
\end{equation}


\newpage

\end{document}